\documentclass[prl,twocolumn,showpacs,preprintnumbers,amsmath,amssymb,superscriptaddress]{revtex4}

\usepackage{graphicx}
\usepackage{dcolumn}
\usepackage{bm}
\usepackage{natbib}
\usepackage{amsmath}

\begin{document}

\title{Tuning non-linear charge transport between integer\\ and fractional quantum Hall states}
\author{Stefano Roddaro}
\affiliation{NEST, Scuola Normale Superiore and CNR-INFM, Piazza dei Cavalieri 7, I-56126 Pisa, Italy}
\author{Nicola Paradiso}
\affiliation{NEST, Scuola Normale Superiore and CNR-INFM, Piazza dei Cavalieri 7, I-56126 Pisa, Italy}
\author{Vittorio Pellegrini}
\affiliation{NEST, Scuola Normale Superiore and CNR-INFM, Piazza dei Cavalieri 7, I-56126 Pisa, Italy}
\author{Giorgio Biasiol}
\affiliation{Laboratorio Nazionale TASC-INFM, Area Science Park Basovizza, I-34012 Trieste, }
\author{Lucia Sorba}
\affiliation{NEST, Scuola Normale Superiore and CNR-INFM, Piazza dei Cavalieri 7, I-56126 Pisa, Italy}
\affiliation{Laboratorio Nazionale TASC-INFM, Area Science Park Basovizza, I-34012 Trieste, }
\author{Fabio Beltram}
\affiliation{NEST, Scuola Normale Superiore and CNR-INFM, Piazza dei Cavalieri 7, I-56126 Pisa, Italy}

\date{\today}

\begin{abstract}

Controllable point junctions between different quantum Hall phases are a necessary building block for the development of mesoscopic circuits based on fractionally-charged quasiparticles. We demonstrate how particle-hole duality can be exploited to realize such point-contact junctions. We show an implementation for the case filling factors $\nu=1$ and $\nu^*\le1$ in which both the fractional filling $\nu^*$ and the coupling strength can be finely and independently tuned. A peculiar crossover from insulating to conducting behavior as $\nu^*$ goes from 1/3 to 1 is observed. These results highlight the key role played on inter-edge tunneling by local charge depletion at the point contact.

\end{abstract}

\maketitle

Fractional quantum Hall (QH) states arise in high-quality two-dimensional systems under the influence of quantizing magnetic fields $B$ as a striking consequence of electron-electron interactions ~\cite{BookQH}. Depending on the fractional value of the filling factor, i.e. the ratio $\nu=n/n_\phi$ between the density of charges $n$ and the density of magnetic-flux quanta $n_\phi=eB/h$, the electron system can condense into a variety of strongly-correlated quantum liquids characterized by an excitation gap in the bulk and exotic fractionally-charged quasiparticles~\cite{QHBase}. Thanks to their unique properties, the latter are interesting candidates for novel mesoscopic circuits and QH-based topological qubits~\cite{TopoQC,TopoQCRec,CM1,CM2}. Additionally, several theoretical articles addressed Andreev-like conversion phenomena involving electrons and fractional quasiparticles at the interface between integer and fractional QH phases~\cite{Fradkin,QHQPCRec}: even device schemes based on these effects were proposed~\cite{Halperin}. One crucial building block of these proposals is the point junction between two distinct QH phases. Its experimental realization, however, poses several challenging problems and for this reason the phenomena described above have been largely unexplored so far in experiments. 

\begin{figure}[h!]
\begin{center}
\includegraphics[width=0.48\textwidth]{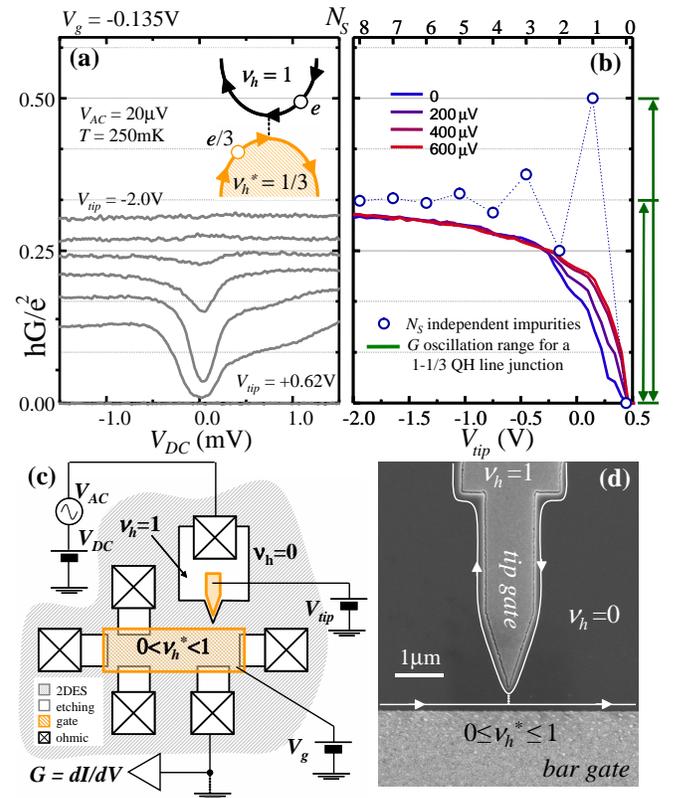}
\caption{(a) Finite bias transport spectroscopy through a point junction between a quantum Hall state at $\nu_h=1/3$ and one at $\nu_h=1$. (b) Comparison between conduction curves at bias $V_{DC}=0$, $200$, $400$ and $600\,{\rm \mu V}$ and theoretical predictions for two selected limit cases: a discrete system of $N_S$ (top axis) uncorrelated tunneling points in the strong coupling regime (hollow blue circles); range of $G$ oscillation for a QH line junction as a function of the interaction length $L$ (green bars on the right side, $G=0$ to $e^2/2h$ or $G=0$ to $e^2/3h$ depending on boundary conditions). (c) Sketch of the device structure and (d) scanning electron microscope picture of the junction region.}
\end{center}
\end{figure}

In this Letter we exploit the concept of particle-hole duality~\cite{Sym1,Sym2} to realize an integer-to-fractional QH point junctions in which the fractional filling-factor value and the interaction strength can be independently tuned. Tunneling conductance measurements are reported that show an insulating behavior when we couple $\nu^*=1/3$ QH edges with integer $\nu=1$ ones; on the other hand, junctions gradually develop a conducting behavior as $\nu^*$ approaches 1. While the insulating behavior is in qualitative agreement with the expected correlation-induced suppression of forward tunneling for fractional edge channels, the observed evolution vs. $\nu^*$  highlights the peculiar role played by the local filling factor value determined by depletion in the junction region between the $\nu$ and $\nu^*$ regions. 

The design of our device is based on particle-hole duality ~\cite{Sym1,Sym2}, which is an exact symmetry of the QH system as long as only the first spin-polarized Landau level is occupied and inter-level mixing can be neglected~\cite{Note}. Duality establishes the occurrence of QH phases for symmetric values of the electron or hole filling factors $\nu_e$ and $\nu_h$ following the mapping $\nu_e \rightarrow\nu_h\equiv 1-\nu_e$~\cite{Mani94,Mani97}. Figure~1c shows a sketch of the device structure, with a more detailed view of the active region shown in the scanning electron micrograph of Fig.~1d. Samples were fabricated starting from a high-mobility AlGaAs/GaAs heterojunction containing a $100\,{\rm nm}$-deep two-dimensional electron system (2DES) with mobility $\mu=1.73\times10^{6}\,{\rm cm^2/Vs}$ and carrier density $n=2.55\times 10^{11}\,{\rm cm^{-2}}$ at $4.2\,{\rm K}$. Ohmic contacts were obtained by evaporation and thermal annealing of a standard Ni/AuGe/Ni/Au multilayer ($5/200/5/100\,{\rm nm}$). All gates were fabricated using an Al/Au bilayer ($10/30\,{\rm nm}$). The main structure consists of a gated region $60\times 930\,{\rm \mu m^2}$ connected to the ohmic contacts by a set of $20\,{\rm \mu m}$-wide etched trenches. When the filling factor in the electron system surrounding the gate is set to $\nu=1$, this multiply-connected circuit realizes a dual version of a standard Hall bar system with a variable filling factor. It is in fact easy to verify that the complementary filling factor configuration $\nu(x)\to\nu_h(x)\equiv1-\nu(x)$, i.e. the filling factor geometry seen from a ``hole perspective'', is exactly the one of a Hall bar at $\nu_h=1$ whose central section can be independently tuned in the range $0\le\nu^*_h\le1$ by adjusting the gate voltage $V_g$. As expected by particle-hole duality, the magneto-transport behavior of this circuit matches the one of a standard electron Hall bar (see supplementary materials). Point junctions were obtained by a further e-beam fabrication step starting from the device structure described above. One of the ohmic contacts was connected to the gated region using a tip-shaped etched structure in place of a macroscopic trench: the tip had an aperture of about $35$ degrees and was fabricated at a nominal distance of $400\,{\rm nm}$ ($426\,{\rm nm}$ measured) off the gate edge. As suggested by the overlay in Fig.~1d, such a structure is meant to induce a point contact between the integer edge propagating around the etched tip and the tunable fractional edge propagating along the gated region. In order to control the exact spatial position of the top edge channel we also fabricated a self-aligned gate metallization inside the etched tip: this electrode allows us to control the spatial extension of the depletion region around the etched tip by adjusting the voltage bias $V_{tip}$, and thus to tune the interaction strength between top and bottom edge channels in Fig.~1d. Thanks to the duality mapping $\nu(x)\to\nu_h(x)\equiv1-\nu(x)$, this device structure can also be fruitfully thought of as an enhancement-mode point contact between a QH liquid at $\nu=1$ and a tunable one at $\nu^*(V_g)$ and thus implements the experimental configuration analyzed in several theoretical works~\cite{Fradkin,QHQPCRec}. Keeping in mind this picture in the following we shall drop the $h$ substript and imply that all filling factors refer to the fraction of {\em unoccupied} states in the first Landau level. While our devices can still be analyzed in terms of electron filling factors and edge systems, the adoption of a hole picture offers a simpler - though equivalent - description of the physical processes at the junction.

Magnetotransport measurements were carried out using a two-wire phase-locked technique at finite bias using a $20\,{\rm \mu V}$ excitation at $17\,{\rm Hz}$. Experiments were carried out in a $^3$He closed-cycle cryostat with $250\,{\rm mK}$ base temperature. The working point for the magnetic field was set in order to configure the electron gas surrounding the device in Fig.~1c to $\nu=1$ at $B=10.0\,{\rm T}$. At this field value the impact of the ohmic-contact resistance was small ($\approx 1\%$) and thus the partition effect on our two-wire junction biasing scheme was negligible and will not be taken into account in the following discussion.

A first relevant transport regime is shown in Fig.~1a, which reports a sequence of differential conductance curves $G(V_{DC})$ at $\nu^*=1/3$ ($V_g=-0.135\,{\rm V}$) for $V_{tip}=-2.00$, $-0.64$, $-0.20$, $+0.10$, $+0.28$, $+0.40$, $+0.64\,{\rm V}$. The point junction in this case connects a fractional QH state at filling factor $1/3$ and an integer one at $\nu=1$. Charge transport is characterized by an insulating behavior as shown by the marked conductance suppression at low bias: this transport fingerprint is consistent with the predicted suppression of forward tunneling stemming from correlations in the fractional edge channel~\cite{Wen,Grayson}. 

Figure~1b compares the $G(V_{tip})$ curves at $V_{DC}=0$, $200$, $400$ and $600\,{\rm \mu V}$ with the predicted conductance $G$ based on two different models of the junction: hollow circles describe $G$ for a set of $N_S$ uncorrelated strongly-coupled point junctions~\cite{Fradkin} while green bars on the right side indicate the range of oscillation of $G$ for a QH line junction as a function of its length and for two different boundary conditions~\cite{QHLJ1,QHLJ2}. The tunable coupling of our device (controlled by $V_{tip}$) allows us to explore all values between the two extremes $G=0$ and $G=e^2/3h$. These transport regimes are obtained for $V_{tip}=+0.75\,{\rm V}$ and $V_{tip}=-2.25\,{\rm V}$, respectively. Note that no overshoot $G>e^2/3h$ is ever observed, contrary to theoretical models for the case of few independent, strongly-coupled tunneling centers~\cite{Fradkin}: these latter studies indicate for instance that in the case of a single tunneling center conductance can exceed the macroscopic junction value $G=e^2/3h$ and be as high as $G=e^2/2h$. In addition, given the small dimension (of the order of $100\,{\rm nm}$) of the interface between the two QH regions, coherence between tunneling centers and Coulomb interactions are likely to play a significant role in our devices. Results in this coherent regime also suggest that G should oscillate strongly as a function of the junction length $L$ (whose value in our devices cannot be determined exactly, but depends on $V_{tip}$) in the range $0\le G\le e^2/2h$~\cite{QHLJ1} or $0\le G\le e^2/3h$~\cite{QHLJ2}, depending on the adopted boundary conditions; no significant damping of the oscillations should be observed as long as the line junction remains coherent, which again is in contrast to what is observed here. The different observed behavior might be linked to the presence of a finite (albeit probably small, given the reduced spatial extension of the junction region of $\approx 400\,{\rm nm}$ and the high-mobility of the 2DES) number of scattering centers with coupling strength increasing from zero up to full-coupling within the explored $V_{tip}$ range. Most importantly, as argued in the following, our work shows that a central role is played by the local filling factor reduction in the constriction region, an experimental feature which is normally not included in point-impurity models.

\begin{figure}[b]
\begin{center}
\includegraphics[width=0.48\textwidth]{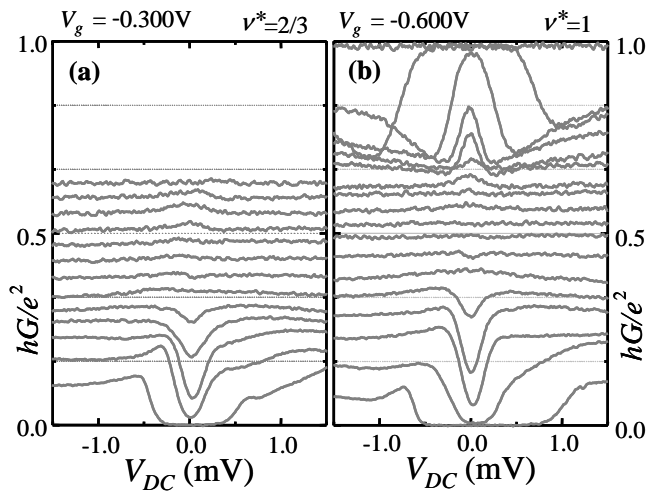}
\caption{Junction transport spectroscopy for two different configurations of $\nu^*=2/3$ (a) and $\nu^*=1$ (b). Different characteristic zero bias anomalies become available when the gate filling factor $\nu^*$ is scaled up to $1$.}
\end{center}
\end{figure}

It is very instructive to tune $\nu^*$ at a given coupling strength. The behavior of the conductance curves reported in Fig.~2a and 2b obtained at the different filling-factor values $\nu^*=2/3$ ($V_g=-0.30\,{\rm V}$) and $\nu^*=1$ ($V_g=-0.60\,{\rm V}$) is revealing. When $\nu^*$ is increased to $2/3$ an additional regime of approximately linear conductance (constant $G$) becomes available in the $e^2/3h < G < 2e^2/3h$ range. As the gate electrode is further biased to reach $\nu^* = 1$ for $V_g=-0.60\,{\rm V}$, a new transport regime characterized by marked positive zero bias anomalies is obtained for $G>2e^2/3h$, in agreement with previous results for constrictions on a QH state at $\nu=1$~\cite{Roddaro3}. The fact that different transport fingerprints become incrementally available while $\nu^*$ is scaled up from $1/3$ to $2/3$ and finally to $1$ clearly indicates the dominant role played by the local junction filling factor $\nu^*_j$ that characterizes the depleted region at the interface between the two QH states. Indeed, while at $\nu^*=1/3$ the possible values for $\nu^*_j$ fall in the $\left[0,1/3\right]$ interval, the limit value of $\nu^*_j$ increases with $\nu^*$: if the junction behavior is determined by $\nu^*_j$, than it is clear that an increase in $\nu^*$ will make new transport regimes accessible.

In order to shed additional light on the observed non-linear transport characteristics, we compared the temperature dependence of the tunneling conductance with that of the QH states in the 2DES region below the macroscopic gate. Figure~3a reports a set of two-wire measurements through the gated region for a junction in the full-coupling regime ($V_{tip}=-2.5\,{\rm V}$): in this configuration $G(V_g,T)$ curves directly sample the conducting properties of the electron liquid below the gate while the junction plays the role of a standard contact. Conductance steps are visible at the base temperature of $250\,{\rm mK}$ for several different QH states under the gate. Quantization values slightly deviate from $\nu^*e^2/h$ (similarly to the values of the saturation conductance observed at $\nu^*=1/3$, $\nu^*=2/3$ and $\nu^*=1$ in Fig.~1a and Fig.~2) owing to residual backscattering phenomena at this relatively high temperature. The temperature evolution in Fig.~3a indicates that bulk fractional QH states melt below $1.0\,{\rm K}$ while the $\nu=1$ QH state survives at higher temperatures. This evolution can be compared in Fig.~3b with the temperature dependence of the non-linear I-Vs obtained for $\nu^*=1$ and $V_{tip}=-0.37\,{\rm V}$. The conductance minimum disappears in the same temperature range of the fractional QH state suggesting that the energy scale of this tunneling conductance is linked not to the $\nu^*=1$ gap but rather to the fractional QH gap associated to $\nu^*_j$. 

\begin{figure}[t]
\begin{center}
\includegraphics[width=0.48\textwidth]{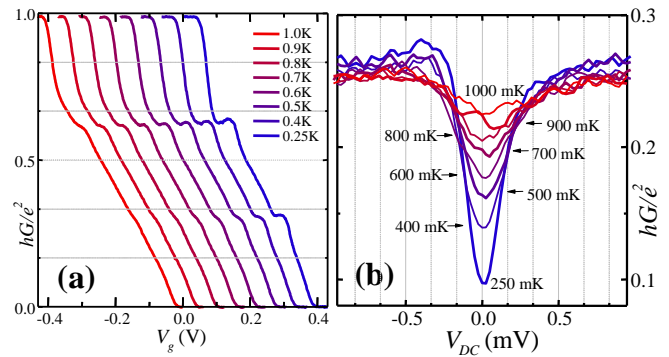}
\caption{Magnetotransport results for the junction transmission for different values of the filling factor $\nu^*(V_g)$. The conductance can be tuned from the saturation value $G=e^2\nu^*/h$ down to zero by changing the bias $V_{tip}$.}
\end{center}
\end{figure}

The role of $\nu^*_j$ in controlling junction behavior raises intriguing questions on the structure of QH edge states at the smooth edges defined by our nanofabricated point-contact junction. Indeed we can argue from the evolution observed in Figs.~1 and 2 that the junction acts as a sort of ``mode selector'' for edge states. Within this interpretative framework, the edge $\nu=1$ can or can not ``branch'' at the junction depending on $\nu^*_j$, with a mechanism that mimics what happens for similar integer QH systems ($\nu=2$ and $\nu^*=1$, for example). While this kind of mechanism is well-established for devices working in the integer QH regime, its extension to the fractional regime appears controversial and no well-established theoretical models are available. We believe that the present experimental results need be taken into account to develop accurate models of the microscopic mechanisms responsible for the non-linear behavior of point-contact junctions in the QH regime~\cite{Roddaro1,Roddaro2,Roddaro3,NonLinearWeizman}, as also suggested by recent theoretical analysis~\cite{HalperinCM,Sidd}. An improved understanding of transport through these QH junctions will also be crucial in relation to the high-contrast interference phenomena in Mach-Zehnder interferometers~\cite{MZ} and to the recent observation of non-linear transport curves - similar to the ones here presented - observed in constrictions on QH states at $\nu=5/2$~\cite{CM1,CM2}. 

In conclusion, we demonstrated a fully-tunable point junction between two different QH states at $\nu=1$ and $\nu^*\le1$. Our fabrication scheme exploits particle-hole duality to successfully implement point junctions. Our transport results showed the significant role of local charge depletion in determining the non-linear junction behavior associated to inter-edge scattering and highlight the need of including this contribution in existing theoretical models.

We thank E.~V. Sukhorukov, P. Roche and C. Strunk for stimulating discussions. We acknowledge support from the Italian Ministry of Research (project RBIN045MNB).


\begin{thebibliography}{9}

\bibitem{BookQH} S. Das Sarma, A. Pinczuk. {\em Perspectives in Quantum Hall Effects} (Wiley, New York, 1997).

\bibitem{QHBase} F.~D.~M. Haldane, J. Phys. C {\bf 14}, 2585 (1981).

\bibitem{TopoQCRec} M.~R. Peterson, T. Jolicoeur, S. Das Sarma, Phys. Rev. Lett. {\bf 101}, 016807 (2008). 

\bibitem{TopoQC} S. Das Sarma, M. Freedman and C. Nayak, Phys. Rev. Lett. {\bf 94}, 166802 (2005).

\bibitem{CM1} I.~P. Radu, J.~B. Miller, C.~M. Marcus, M.~A. Kastner, L.~N. Pfeiffer, K.~W. West, Science {\bf 320}, 899 (2008).

\bibitem{CM2} M. Dolev, M. Heiblum, V. Umansky, A. Stern, D. Mahaluu, Nature {\bf 452}, 829 (2008).

\bibitem{Fradkin} C.~C. Chamon and E.Fradkin, Phys. Rev. B {\bf 56}, 2012 (1997).

\bibitem{QHQPCRec} V.~V. Ponomarenko and D.~V. Averin, Phys. Rev. B {\bf 70}, 195316 (2004).

\bibitem{Halperin} D.~B. Chklovskii and B.~I. Halperin, Phys. Rev. B {\bf 57}, 3781 (1998).

\bibitem{Sym1} S.~M. Girvin, Phys. Rev. B {\bf 29}, 6012 (1984).

\bibitem{Sym2} A.~H. MacDonald and E.~H. Rezayi,
Phys. Rev. B {\bf 42}, 3224 (1990).

\bibitem{Note} This condition is satisfied in the limit of high magnetic fields. For this reason all data shown in the paper were taken at a magnetic field of $10\,{\rm T}$, which is well above known spin transitions for fractional QH states studied here $\nu=1/3$ and $\nu=2/3$.

\bibitem{Mani94} R.~G. Mani and K. von Klitzing, Appl. Phys. Lett. {\bf 64}, 1262 (1994).

\bibitem{Mani97} R.~G. Mani, Phys. Rev. B {\bf 55} 15838 (1997).

\bibitem{Wen} X.G. Wen, Phys. Rev. B {\bf 41}, 12838 (1990).

\bibitem{Grayson} M. Grayson, D.~C. Tsui, L.~N. Pfeiffer, K.~W. West, A.~M. Chang, Phys. Rev. Lett. {\bf 80}, 1062 (1998).

\bibitem{QHLJ1} V.~V. Ponomarenko and D.~V. Averin, Phys. Rev. Lett. {\bf 97}, 159701 (2006).

\bibitem{QHLJ2} U. Z\"ulicke and E. Shimshoni, Phys. Rev. B {\bf 69}, 085307 (2004).

\bibitem{Roddaro1} S. Roddaro, V. Pellegrini, F. Beltram, G. Biasiol, L. Sorba, R. Raimondi, G. Vignale, Phys. Rev. Lett. {\bf 90}, 046805 (2003).

\bibitem{Roddaro2} S. Roddaro, V. Pellegrini, F. Beltram, G. Biasiol, L. Sorba, Phys. Rev. Lett. {\bf 93}, 046801 (2004).

\bibitem{Roddaro3} S. Roddaro, V. Pellegrini, F. Beltram, L.~N. Pfeiffer and K.~W. West, Phys. Rev. Lett. {\bf 95}, 156804 (2005).

\bibitem{NonLinearWeizman} Y.~C. Chung, M. Heiblum, V. Umansky, Phys. Rev. Lett. {\bf 91}, 216804 (2003).

\bibitem{HalperinCM} B. Rosenow, B. I. Halperin, cond-mat.mes-hall/0806.0869v2

\bibitem{Sidd} S. Lal, Phys. Rev. B {\bf 77}, 035331 (2008).

\bibitem{MZ} Y. Ji, Y. Chung, D. Sprinzak, M. Heiblum, D. Mahalu, H. Shtrikman, Nature {\bf 422}, 425 (2003).

\end{thebibliography}
\end{document}


\begin{center}
{\bf \huge Supplementary Material}
\end{center}
\vspace{0.3cm}

\begin{center}
{\bf \large Fully-dual quantum Hall bars}
\end{center}

The two precisely dual version of a quantum Hall bar with tunable filling factor were implemented following the mapping $\nu_e(x)\rightarrow\nu_h(x)=1-\nu_e(x)$ between the standard electron filling factor $\nu_e$ and corresponding the hole filling factor $\nu_h$ starting from the filled first Landau level as the system vacuum~\cite{Note}. The resistivity of these two circuits was found to be completely equivalent at a magnetic field of $B=ne/h=10\,{\rm T}$ where $n$ is the carrier density in the two-dimensional electron system used for this research work.
\newline

A standard Hall bar is shown in Fig.~1a: the mesa is defined by chemical etching and carrier density in the central part of the conducting channel can be controlled by field effect thanks to a Schottky top gate. When the magnetic field $B$ is set to $B=nh/e$ a filling factor $\nu_e=1$ is obtained in the ungated region while the filling factor $\nu_e^*$ beneath the gate can be controlled in the $0<\nu_e^*<1$ range by tuning the gate bias $V_g$. Figure~1c shows the evolution of the longitudinal resistance at $10\,{\rm T}$ as a function of the gate bias. Starting from the dissipationless transport at $V_g=0$ due to the presence of an extended quantum Hall state at $\nu_e=1$ between the ohmic contacts, longitudinal resistivity increases when the electron system becomes compressible. Fractional quantum Hall states $\nu_e^*=1/3$ and $\nu_e^*=2/3$ can be clearly identified as strong resistance dips: the relatively high temperature ($250\,{\rm mK}$) at which the measurement was performed however does not allow the observation of a completely dissipationless transport ($\rho_{xx}=0$) for these fractions. For $V_g<-0.40\,{\rm V}$ the electron system beneath the gate becomes completely depleted and resistivity diverges.

\begin{figure}[h!]
\begin{center}
\includegraphics[width=0.8\textwidth]{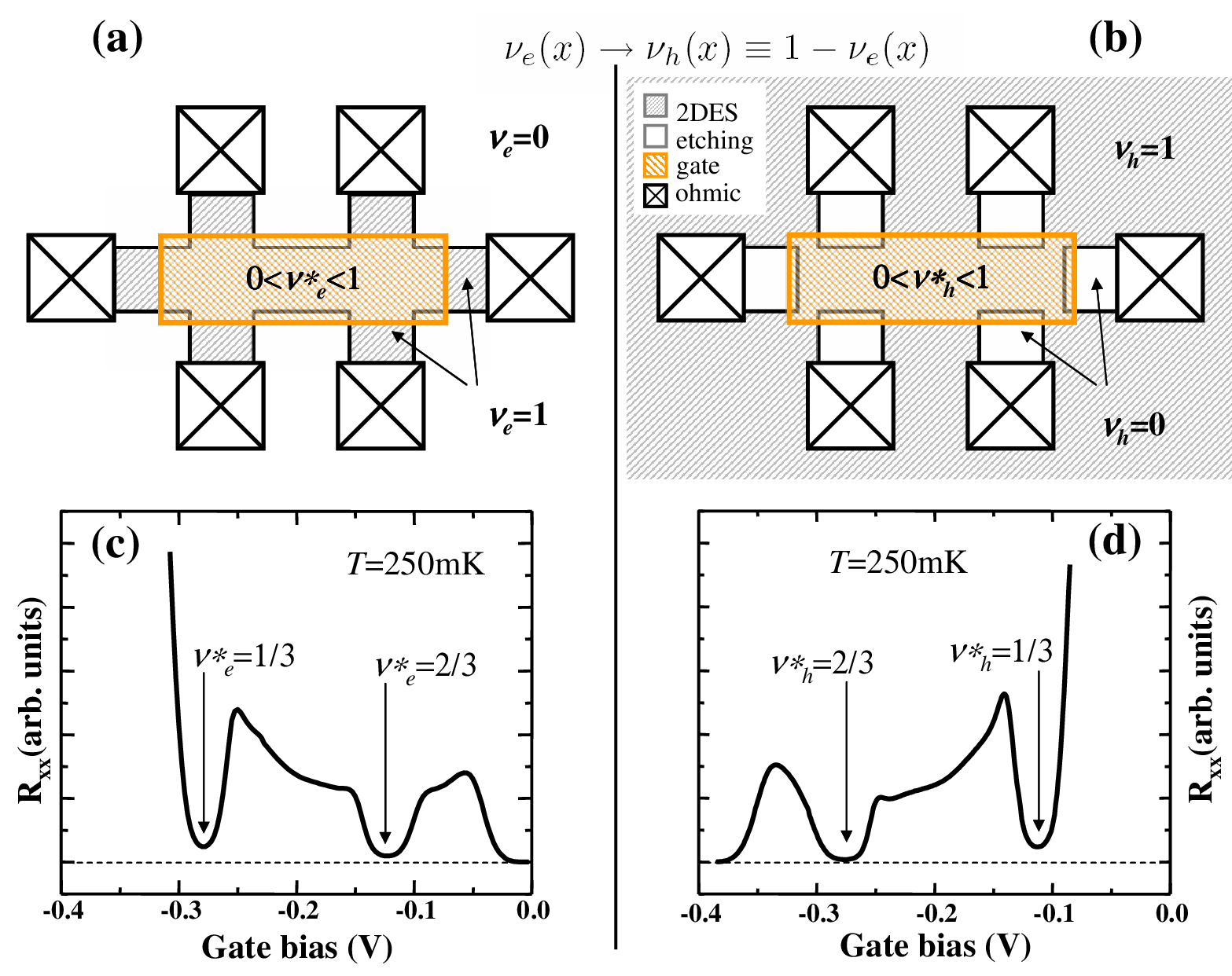}
\caption{Two dual versions of the same circuit are implemented using a standard electron filling factor geometry $\nu_e(x)$ on the left side and hole filling factors $\nu_h(x)=1-\nu_e(x)$ on the right side.}
\end{center}
\end{figure}

Following the duality mapping, the same circuit shown on the left side of Fig.~1a was also implemented in terms of holes as visible in Fig.~1b: the 2DES surrounding the dual Hall bar is intact and set to $\nu_e=1$ at $10\,{\rm T}$; the Hall bar is again defined by a top gate that allows to control the hole density, i.e. the hole filling factor $\nu_h^*(V_g)$; the Hall bar arm are kept at filling $\nu_h=1$, which correponds to complete electron depletion that in our case was obtained by physically removing the heterojunction by chemical etching. Geometrical proportions were preserved in the two circuits. Even if the circuit in Fig.~1b is characterized by a complex multiply-connected geometry and the surroding electron system provides uncontrolled leakage paths at $B=0\,{\rm T}$, the circuit behavior at $10\,{\rm T}$ becomes simple again and is completely similar to the one observed for the standard version of the device. In this case (Fig.~1d) dissipationless transport is observed for $V_g<-0.40\,{\rm V}$, i.e. when the electron system is completely depleted and $\nu^*_h=1$. The gated region becomes compressible for higher bias voltages and this leads to an increase in the sample resistivity. Again, low dissipation transport is obtained when the gated region reaches a filling $\nu^*_h=1/3$ ($\nu^*_e=2/3$) and $\nu_h^*=2/3$ ($\nu^*_e=1/3$). Finally at $V_g=0\,{\rm V}$ the gated region cannot support transport between the different ohmic contacts and resistivity diverges. Note that the relative strength of the two fractional QH dips, as well as the general evolution of the $\rho_{xx}(V_g)$ curve, comply very well with the duality mapping.